\documentclass[12pt, oneside]{article}   	
\usepackage{textcomp}
\usepackage{geometry}                		
\usepackage{graphicx}				
\usepackage{caption}
\usepackage{subcaption}								
\usepackage{color}
\usepackage{amssymb}
\usepackage{amsthm}

\usepackage{natbib}
\usepackage{xcolor}
\usepackage{authblk}
\usepackage{float}
\usepackage{rotating}
\usepackage{adjustbox}
\usepackage[font=small,labelfont=bf]{caption}
\usepackage[final]{changes}
\definechangesauthor[name={George Chacko}, color=blue]{gc}

\usepackage{authblk}
\title{AOC: Assembling Overlapping Communities}
\author[1]{Akhil Jakatdar\thanks{AJ and BL contributed equally to this study.}}
\author[1]{Baqiao Liu}
\author[1]{Tandy Warnow\thanks{warnow@illinois.edu}}
\author[1,2]{George Chacko\thanks{chackoge@illinois.edu}}
\affil[1]{Department of Computer Science, University of Illinois Urbana-Champaign, Urbana, IL 61801}
\affil[2]{Office of Research, Grainger College of Engineering, University of Illinois Urbana-Champaign, Urbana, IL 61801}

\begin{document}
\maketitle
	
\abstract{Through discovery of meso-scale structures, community detection methods contribute to the understanding of complex networks. Many community finding methods, however, rely on disjoint clustering techniques, in which node membership is restricted to one community or cluster. This strict requirement limits the ability to inclusively describe communities since some nodes may reasonably be assigned to many communities. We have previously reported Iterative K-core Clustering (IKC), a scalable and modular pipeline that discovers disjoint research communities from the scientific literature. We now present Assembling Overlapping Clusters (AOC), a complementary meta-method for overlapping communities as an option that addresses the disjoint clustering problem. We present findings from the use of AOC on a network of over 13 million nodes that captures recent research in the very rapidly growing field of extracellular vesicles in biology.
}
	
\clearpage
	
\section{Introduction} 

We are motivated by the problem of identifying and characterizing research communities in the network of scientific activity. Research communities represent scientific specialization \citep{Chubin1976,Morris2009} that evolves in response to influences such as new research paradigms, policy, collaboration practices, and increasing globalization. We are interested in scalable community detection methods for identifying research communities as they emerge and grow to maturity. We would also like to understand the extent to which these communities overlap. 

A community in a network generally refers to a group of nodes that are more densely connected with each other than to nodes outside the community. Various flavors of this definition exist, and the terms community and cluster have overlapping uses  \citep{Coscia2011,yang2013overlapping}. 
  
A considerable literature exists on community finding in graphs that reflects a diversity of perspectives and solutions. Hence various approaches can be drawn upon in identifying  communities, of which we cite a few: \cite{Fortunato2009,Fortunato2010,Coscia2011,Yang2016}. Most community finding approaches focus on disjoint partitioning, where a vertex, or node, is only assigned to only one community. This restriction can be viewed as a limitation since some nodes can reasonably be assigned to more than one community.

Several methods have been developed that address the limitation of disjoint clustering by producing overlapping clusters \citep{Baumes2005,Palla2005,banerjee2005model,Cleuziou2008,Lancichinetti2009,Lu2012,yang2013overlapping},  although these do not appear to enjoy much use in scientometric studies. 

An interesting two-step procedure has been proposed to enable overlapping clusters: the input graph is first transformed into a line graph whose nodes represent edges in the original graph \citep{Harary1960}.  The line graph is then clustered and this output can be mapped back to the input graph to generate overlapping clusters. This general approach has been used by others on citation graphs \citep{Evans2009,Havemann2018,Havemann2021}. However, line graph  techniques are not very scalable, since the size of a line graph is much larger than the size of its input graph. For example, the network studied in \cite{Wedell2022} would grow from 13,989,436 nodes and 92,051,051 edges to 92,051,051 nodes and 160,428,881,121 edges, which presents a challenge to clustering software. 
 
For the specific problem of identifying research communities, one approach is through analyzing citation patterns in the scientific literature. The underlying assumption is that publications in a research community are more likely to cite each other's work than the work from outside their community. Drawing upon the rich literature in graph theory, this question can be framed as a community finding problem where a community is defined as a set of vertices in a graph that exhibit stronger connectivity to each other than to vertices outside such a community. Thus, in the graph (or network) of scientific literature, citation-dense areas suggest the existence of communities of publications. Accordingly, a community of publications can be defined by edge-density and its researcher members can then be inferred from the authorship of these publications \citep{Chandrasekharan2021,Wedell2022}.

Community finding and clustering approaches have been applied to the scientific literature~\citep{Newman2006,Fortunato2009,Boyack2010,Boyack2019,Traag2019,Ahlgren2020,Chandrasekharan2021,Wedell2022}. As in other application areas, disjoint methods are also limited when studying citation networks \added[id=gc]{since} publications introducing widely used methods may be relevant to multiple communities.
	
Beyond identifying research communities, we are interested in their structure and the roles played by community members.  The observation by \cite{Price1966} that a community of oxidative phosphorylation researchers consisted of a small core of influential researchers and a much larger transient population drew attention to core-periphery or center-periphery structure in research communities. Core-periphery patterns have also been reported in other networks using different techniques, such as block modeling and k-core decomposition, arguing for some degree of ubiquity in their occurrence \citep{borgatti2000models,Breiger2014,Zhang2015,Rombach2017,gallagher2021clarified,yanchenko_2202.04455}. 	

We recently reported Iterative K-core Clustering or IKC \citep{Wedell2022}, a recursive algorithmic approach based on the k-core property \citep{Giatsidis2011,malliaros2019}, which helps identify densely connected parts of a graph--the cores of core-periphery structures. Specifically, a k-core in a graph is a maximal connected subgraph where every node in the subgraph has at least k neighbors in the graph.
The largest value for k for which a k-core exists in a graph is called its ``degeneracy". IKC recursively extracts disjoint k-cores from a graph beginning with the largest value of k for which a  k-core  exists, and then reducing k until some user-specified lower bound on k is reached. IKC returns those clusters it finds that have positive modularity.  \added[id=gc]{However, IKC does not enforce global modularity maximization \citep{lancichinetti2011limits}, which suffers from the theoretical problem of the resolution limit that favors larger clusters that can contains smaller clusters within \citep{Fortunato2006}. Instead we require only that each cluster exhibit positive modularity, a mild constraint on cluster quality}.

We implemented IKC as the first step in a tunable modular pipeline to identify communities with core-periphery structure. 
Subsequent steps in the pipeline include breaking large cores and adding peripheral nodes to each core to construct communities with core-periphery structure. While IKC is a necessary step in the pipeline, the remaining steps are optional. We tested the ability of IKC to recursively extract k-cores from a network where nodes were publications and edges were citations. 
The input was a network of greater than 14 million articles centered around the rapidly growing field of extracellular vesicle biology \citep{Wedell2022}. Using this pipeline, we were able to reduce a large network to two principal communities of interest that were robust to various option settings \citep[Figure 5]{Wedell2022} from this large dataset.
	
However, IKC produces disjoint clusters. As noted above, this limitation impacts articles that describe widely used methods but is also relevant, in theory at least, to articles reporting discovery that are influential in more than one community.  Thus, a need exists for methods that can produce meaningful overlapping clusters. 
	
To address the limitation of disjoint clusters with IKC, we developed  ``Assembling Overlapping Clusters" (AOC), a scalable meta-method that takes the output of IKC and makes multiple community assignments from a list of candidate nodes, while enforcing criteria for cores. AOC can be used as an optional step in the IKC pipeline at the discretion of the user. We present results from AOC applied to the cores generated by the IKC method, and discuss the results and discovery made from them. We use the terms core, community, and cluster interchangeably in this article.

 \added{In this study, as in our prior work \citep{Wedell2022}, we are focused on the  problem of finding  research communities of publications. However, we note that this allows us, in a second pass, to extract the authors of these publication communities, and hence subsequently also obtain author communities.}  We stress that citation density alone does not make a confirming argument for the existence of a research community. However, community finding techniques are valuable in being able to efficiently search large datasets for communities, reducing them to smaller units that can then be examined with complementary analytical techniques that include the use of human judgment. 

\section{Materials and Methods}
	
\subsection{Methods} Motivated by the graph-theoretic concept of k-cores \citep{Giatsidis2011,malliaros2019}, we have previously constructed a clustering pipeline we refer to as  the Iterative K-core Clustering (IKC) pipeline in \cite{Wedell2022}. The input to the IKC pipeline is a network $N$ and the output is a set of disjoint clusters, where each cluster has a ``core'' component, and a ``periphery".  To produce this clustering,
the IKC pipeline takes, as user-selected algorithmic parameters, two positive integers $k$ and $p$ with $k > p$, and computes a clustering of a given network $N$ into disjoint clusters 
This clustering is designed to satisfy several criteria: (i) the core is connected,  has positive modularity ($m$-valid), and each node in the center  is adjacent to at least $k$ other nodes in the center ($k$-valid), and (ii) every node in the periphery of a cluster is adjacent to at least $p$ center nodes in the cluster ($p$-valid). Thus, membership in the core of a cluster requires a greater degree of connectivity to the other center nodes than membership in the periphery. 

The IKC pipeline has three basic steps, where the second and third steps are optional.  The first step (the iterative k-core extraction algorithm) produces disjoint clusters that are both $k$-valid and $m$-valid, i.e.,  $km$-valid, where each cluster has positive modularity and each node in each cluster is adjacent to at least $k$ other nodes in the cluster; these form the centers or cores of the communities. The optional second and third steps breaks these clusters into smaller clusters and adds peripheries to the clusters respectively.  Note that the parameter $k$ is used to define the centers and the parameter $p$ is used to define the periphery. If the only objective is cores or centers of communities, then the pipeline can be run using only the first step (or optionally also with the second step if smaller communities are desired).

The \emph{Assembling Overlapping Clusters (AOC)} method, presented herein, builds on k-core extraction, the first step of the IKC pipeline. The input to AOC is a network $N$ and a set of disjoint clusters produced by the first step of the IKC pipeline. These clusters will be expanded by AOC,  using a simple technique that we now describe. To run AOC, the user specifies additional algorithmic parameters:  (i) the set of candidate nodes that are being considered for membership in the expanded clusters and  (ii) the criterion for adding nodes to the expanding clusters. 
Thus, we refer to the network $N$ and the clustering as the input to AOC, and the set of candidate nodes and criterion for node inclusion in additional clusters as the algorithmic parameters. However, these may also be considered part of the input since the user specifies these values.

The user can specify any  set of candidate nodes, but obvious options are using node-network characteristics, such as  being in selected  input clusters,  total degree, in-degree,  out-degree,  or some other basis for candidate selection such as publication venue or funding sources. The candidate nodes are then sorted by their total degrees in decreasing order (largest degree first), and are processed one-by-one for each input cluster. The criterion for adding node $v$ to the expansion of cluster $C$ (where $C$ is in the input clustering) is selected from the following two criteria:
 (i) $v$ must have at least $MCD(C)$ neighbors in  $C$, 
where $MCD(C)$ is the minimum core degree (i.e., the minimum number of neighbors of any node within $C$)  and
(ii) $v$ must have at least $k$ neighbors in  $C$. We note that $k$ (the parameter for condition (i)) has been used to construct the IKC clustering; hence, for every cluster $C$,  it follows that $MCD(C) \geq k$ and so (ii) is a weaker condition than (i). We refer to the first membership criterion as AOC\_m and the second as AOC\_k.
Independent of the selected membership criterion, we also require that the addition of $v$ to the current expansion of $C$ produce a cluster that has positive modularity.
This last criterion ensures that the final expanded cluster will also have positive modularity.
	Thus, whether running   AOC\_m or AOC\_k on the IKC clustering, both pipelines produce a set of expanded clusters, relative to the input clustering, that are potentially overlapping,  each of which has positive modularity and where every node in every cluster has at least $k$
	 other nodes in the cluster.  
	 	
The resulting clustering generated from the overlapping clusters construction stage can now contain nodes in multiple clusterings, a property not found in the IKC method. 
Since the input clustering is produced by IKC, every  expanded cluster that is produced is still $km$-valid. Furthermore, if AOC\_m is used as the criterion, then the MCD of each cluster is preserved.

The AOC technique we describe can be modified to suit the user's interests. For example,   the user may wish to modify the input IKC clustering through the optional second step, which breaks  up large clusters but still ensures that the clusters are km-valid.
If positive modularity is not required, then the check for modularity could be dropped.  Finally these expanded clusters can also be enlarged through the inclusion of peripheral nodes through Step 3 in the IKC pipeline.

\subsection{Data} 

\emph{Citation network} We previously generated a citation network \citep{Wedell2022} representing the exosome literature and more generally the extracellular vesicle literature \citep{harding1983,raposo2021} from the Dimensions database \citep{hook2018dimensions} in the Google cloud. \added[id=gc]{Briefly, the network was constructed by first performing a lexical search for the term ``exosome'', which was labeled as ``S", the seed set. This seed set was amplified using the SABPQ protocol described in \cite{Wedell2022} to capture articles linked to the seed set by citation and form a final network where each node in the network is a publication and each edge is a direct citation. For the present study, we curated this exosome-centric network to deplete it of both retracted articles and relatively high-referencing articles. Retractions were identified from a database kindly provided by Retraction Watch \citep{rw2022} and matched to nodes in the network using digital object identifiers (DOIs). Any article with 250 or more references was also removed. While the network in \cite{Wedell2022} consisted of 14,695,475 nodes and 99,663,372 edges, the network resulting from removing retracted and high-referencing articles comprised 13,989,436 nodes and 92,051,051 edges. Thus, 706,039 nodes and 7,612,321 edges were removed. We refer to this network as the Curated Exosome Network (CEN). Its largest connected component consists of 13,988,426 nodes and accounts for 99.99\% of the CEN.}
	
\emph{Marker nodes}. To identify exosome-relevant publications and communities, we re-used a set of marker nodes described in  \cite{Wedell2022}. These 1,218 markers are the cited references from 12 different recent review articles on exosomes and extracellular vesicles. Of these, 1,021 are present in the CEN and  were used to identify clusters relevant to extracellular vesicles research.

\emph{Random networks.}
We also explored clustering on random networks, specifically using both Erd\H{o}s-Renyi (ER) graphs and configuration models.  The
ER graphs  analyzed in this paper were generated using the Python package NetworkX \citep{hagberg2008} using the random graph function that requires four parameters: number of input vertices, number of input edges, random seed value (for reproducibility) and a Boolean value for whether the randomly generated graph is directed or not. Using this function, 100 ER graphs were generated using seed values from 0 to 99 inclusive. An example of a command used to generate a single graph is  `nx.gnm\_random\_graph(n=13989436, m=92051051, seed=0, directed=True)'. 
We also constructed configuration null models where the edges of the input network were randomized while preserving the total number of nodes, the degree of each node,  and the publication year of each cited node in a citing-cited node pair \citep{bradley2020}. 
 
\section{Results and Discussion}
	
As we explain in the preceding sections, AOC is a meta-method for overlapping communities that takes as input a network $N$ and a clustering of the network produced by IKC,  and has user-specified parameters: (i) a set of candidate nodes for consideration of membership in multiple communities, and (ii) a parameter $k$ or $m$ that defines the criterion for membership (Materials and Methods). We now examine the properties of non-disjoint clusterings produced using IKC followed by AOC. We analyze the effects of AOC on IKC clusters using either the nodes in non-singleton IKC clusters as candidates (the second input parameter) or high-degree nodes in singleton clusters. 
We also study the distribution of marker nodes in IKC clusters enhanced by AOC and, finally, we examine overlap across AOC clusters. 
	
\subsection{Characterizing the Curated Exosome Network}
 In an initial exploratory experiment, we clustered the curated exosome network (CEN) using IKC where $k \in {\{10,20,30,40, 50\}}$, and we refer to these clusterings as
 IKC\_{k10},  IKC\_{k20}, etc. 
At the value of $k$ with maximum coverage (k=10) in the CEN, 128 $km$-valid cores ($k \geq 10$ and modularity $> 0$) containing a total of 535,165 nodes (3.8\% of the CEN network) are discovered. These cores range in size from 14 to 214,877, with a median core size of 79, and minimum core degree (MCD) varying from 10 to 53 with median MCD of 16. Thus, the CEN is a 53-degenerate graph consisting of 13,989,436 vertices.

The core sizes and MCD values for this curated exosome network are very close to the core sizes and MCD values of the original exosome network (prior to curation) studied in \cite{Wedell2022}.
Specifically, the impact of curation reduced coverage
 from 4.2\% to 3.8\%, reduced the highest MCD value among the cores from 56 to 53, reduced the median core size from 85 to 79, and increased the number of clusters  from 119 to 128.
 Thus, the benefit of curation, which removed retracted and high-referencing articles, does not come at a steep cost.

\begin{figure}[H]
\centering
\includegraphics[width=0.6\linewidth]{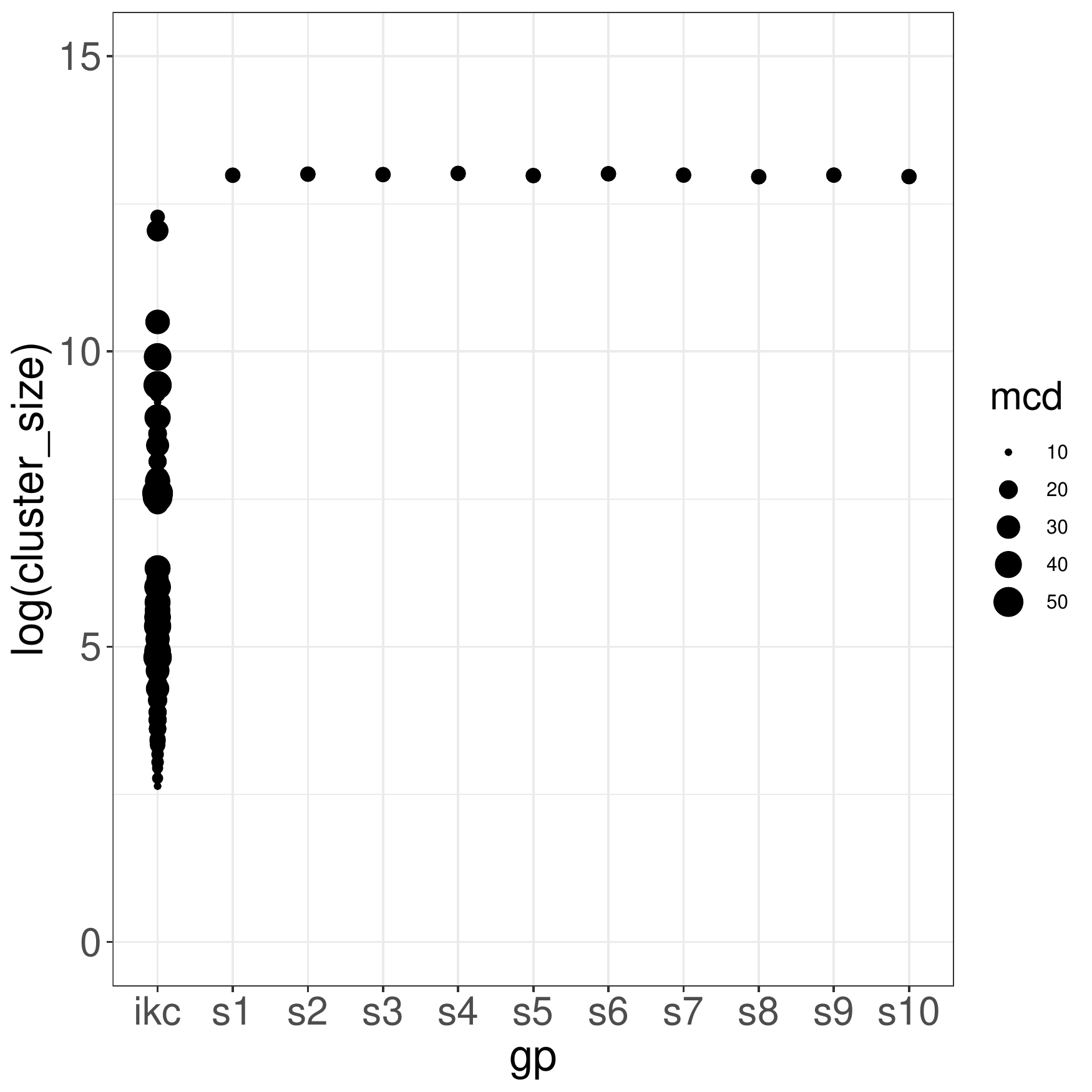} 
\captionsetup{width=0.9\textwidth}
\caption{IKC clustering of Configuration Null Model. IKC clustering with k=10 of the original CEN network produced 128 clusters  with MCD ranging from 10 to 52. The edges of the CEN network were randomly shuffled while preserving degree distribution for each node and the year of publication for citing and cited nodes. The resultant networks were clustered with IKC with k=10 (IKC\_k10).  In all 10 cases, a single k-core with MCD=15 resulted, although the size of this core varied slightly between replicates (s1, s2, ..., s10). Cluster size is shown on the y-axis in natural log units. Log values of 5,10, and 15 correspond, after rounding, to 148, 22,026, and 3,269,017 respectively.}
\label{fig:shuffleplot}
\end{figure}

To examine random network effects that could contribute to observed results from IKC, we generated 10 replicates of a configuration null model where the edges of the input network were randomized while preserving the total number of nodes, the degree of each node,  and the publication year of each cited node in a citing-cited node pair \citep{bradley2020,uzzi2013}. These shuffled networks were then clustered using IKC\_k10. In all 10 cases, only a single $km$-valid core with MCD of 15 was extracted, although the number of nodes in each of these cores varied (Figure~\ref{fig:shuffleplot}).
 The median size of this cluster across 10 replicates was 435,216, roughly double the size of the largest cluster generated from the unperturbed network. These observations suggest that under these controlled conditions of randomization, the collective affinity of community members expressed as internal edge density is disrupted and results in a single k-core. We did not run IKC at higher values of $k$ on the shuffled networks since  by definition no clusters would have been found. We also did not run AOC on the output from IKC with  k=10 either,  since AOC cannot add new members when the output has only one cluster and the candidate node list is restricted to membership of that cluster.

We also generated 100 random Erd\H{o}s-Renyi graphs with the same number of nodes and edges as the CEN network and clustered them with IKC with $k=10$ (Supplementary Material). No $km$-valid clusters were generated from these Erd\H{o}s-Renyi graphs and in particular, we found that the degeneracy of each of the Erd\H{o}s-Renyi graphs was 9. Because there were no clusters in the IKC output,  it was meaningless to run AOC.

The distinct differences in results from the real world CEN network and the two random network models we explored show that the results seen in IKC clustering on a real world network are unlikely to be the result of random effects alone. 
	
	
	
\subsection{Effect of AOC on IKC clusters} 
As we assert, a limitation of disjoint clustering methods is that restricting membership to one community excludes assignment to other communities where a node may have both role and influence. Since clustering with IKC occurs iteratively with the densest core being extracted first, nodes in an extracted core are not considered for membership in cores that are subsequently extracted.

Accordingly we asked whether AOC could add nodes from disjoint cores generated by IKC to other cores in the same clustering. Note that we refer to the clusters produced by IKC as ``cores"; this is because they form the centers or cores within the center-periphery clusterings we produced using the full IKC pipeline, of which the IKC component is just the first step. In this experiment,  we set the algorithmic parameters for AOC as follows: (i) the clustering produced by IKC with the CEN network was used as input to AOC with $k \in \{10, 20, 30, 40, 50\}$; (ii) the set of candidate nodes was every node within the cores generated by IKC; and (iii) the criterion for membership was either (i) AOC\_m or AOC\_k (defined above).

\begin{table}[H]
\centering
\begin{tabular}{rlcc}
\hline
& AOC\_m & \# clusters that do not change & \# clusters that increase\\ 
\hline
 1 & ikc10 &  33 &  95 \\ 
  2 & ikc20 &  10 &  34 \\ 
  3 & ikc30 &   8 &  14 \\ 
  4 & ikc40 &   3 &   3 \\ 
  5 & ikc50 &   1 &   0 \\ \hline
\end{tabular}
\quad
\begin{tabular}{rlcc}
\hline
& AOC\_k& \# clusters that do not change & \# clusters that increase \\ 
  \hline
 1 & ikc10 &  17 & 111 \\ 
  2 & ikc20 &   2 &  42 \\ 
  3 & ikc30 &   4 &  18 \\ 
  4 & ikc40 &   2 &   4 \\ 
  5 & ikc50 &   1 &   0 \\ 
   \hline
\end{tabular}
\caption{The number of clusters that change or do not change in size, after AOC\_m or AOC\_k treatment.}
\label{tab:tab-change-no-change}
\end{table}

By construction, the number of cores cannot change by running AOC  under any setting of its algorithmic parameters. However, core sizes can increase, with increases resulting from AOC\_k being at least as large as increases resulting from AOC\_m. Of interest, therefore, is how the algorithmic parameters, such as the value for $k$ and the specified subset of nodes, impact the increase in size, and how node properties, such as degree, influence the number of clusters they are added to.
	
\begin{figure}[H]
\centering
\begin{subfigure}[t]{0.48\textwidth}
\centering
\includegraphics[width=\linewidth]{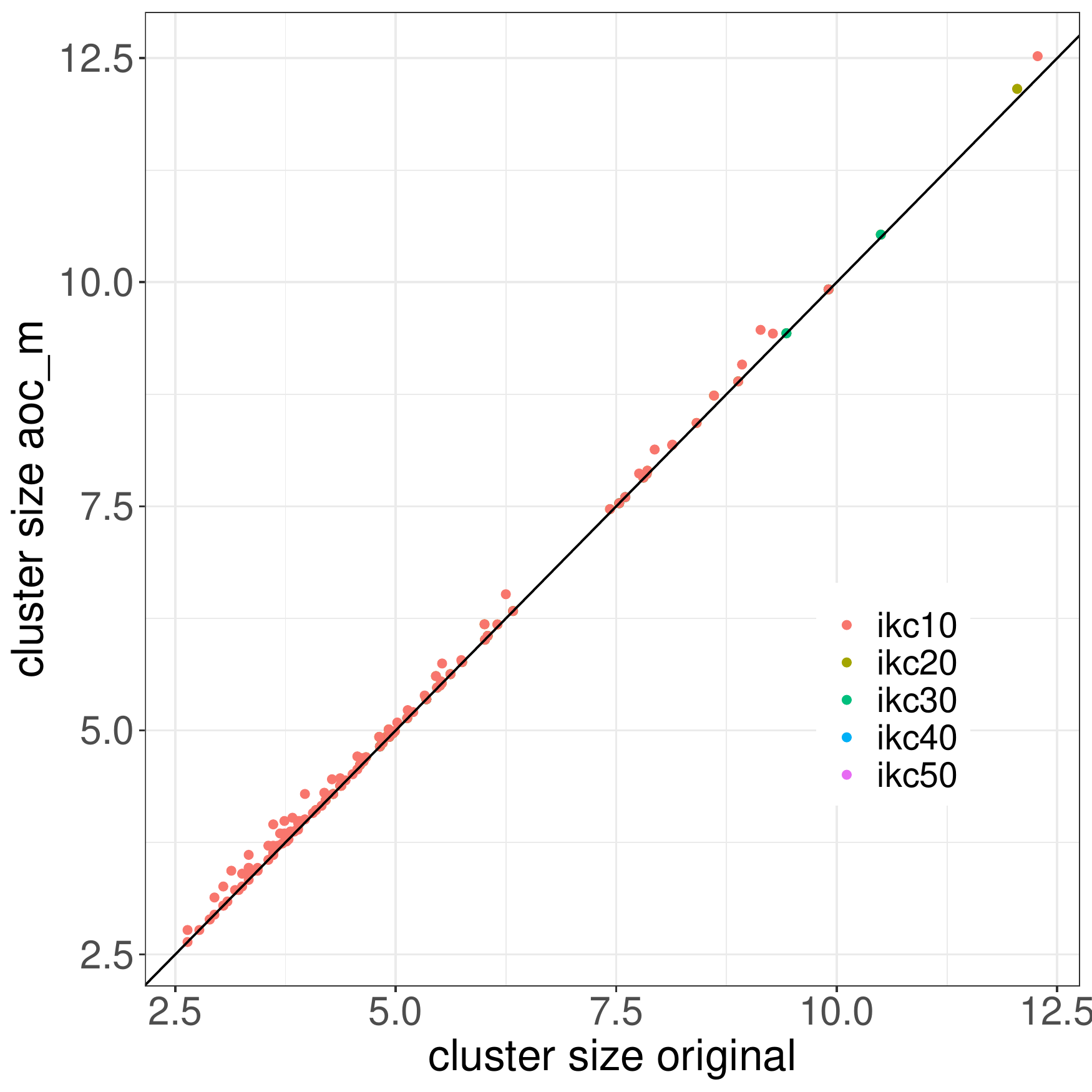} 
\end{subfigure}
\hfill
\begin{subfigure}[t]{0.48\textwidth}
\centering
\includegraphics[width=\linewidth]{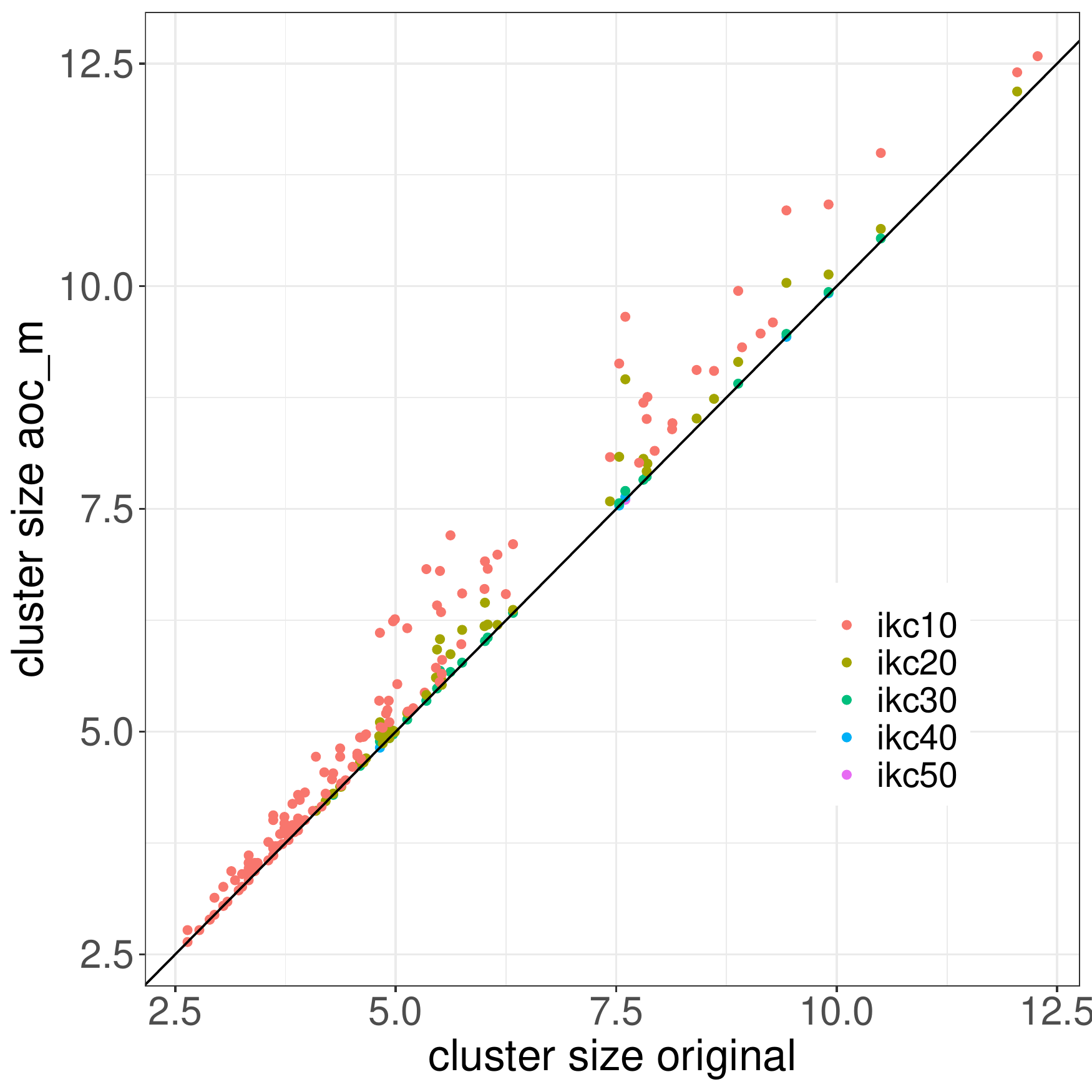} 
\end{subfigure}
\captionsetup{width=0.9\textwidth}	
\caption{Comparison of cluster sizes between disjoint (IKC)  and overlapping (AOC) clusters. Clusters were generated from the CEN network by IKC using values of $k$ ranging from 10 to 50. These clusters were then enriched through the AOC process enforcing either \emph{mcd} (left panel) or \emph{k} (right panel). The input to AOC was the  clustering produced by 
IKC and the set of candidate nodes to be assigned additional clusters was all nodes in non-singleton IKC clusters.  Points that lie on the diagonal indicate no change in cluster size after AOC treatment. A natural log scale is used for both axes. Log values of 2.5, 7.5, and 12.5 correspond, after rounding, to 12, 1808, and 268,337 respectively.}
\label{fig:fig1}
\end{figure}

For both AOC\_m and AOC\_k a subset of the clusters increases in size (Table~\ref{tab:tab-change-no-change}). 
Figure~\ref{fig:fig1} shows the distribution of cluster sizes generated by AOC relative to the core sizes from the IKC run at various values for \emph{k}.  Approximately 74\% and 87\% of  128 cores increase in size with AOC\_m and AOC\_k treatment respectively when IKC with $k$=10 is used as the input clustering. AOC treatment of IKC clustering, therefore, results in an increase in cluster sizes that is inversely related to the value of \emph{k} used in IKC and is more pronounced with AOC\_k than with AOC\_m. For reasons of coverage, we used IKC with $k$=10 in all subsequent experiments.
	
\begin{figure}[H]
\centering
\begin{subfigure}[t]{0.48\textwidth}
\centering
\includegraphics[width=\linewidth]{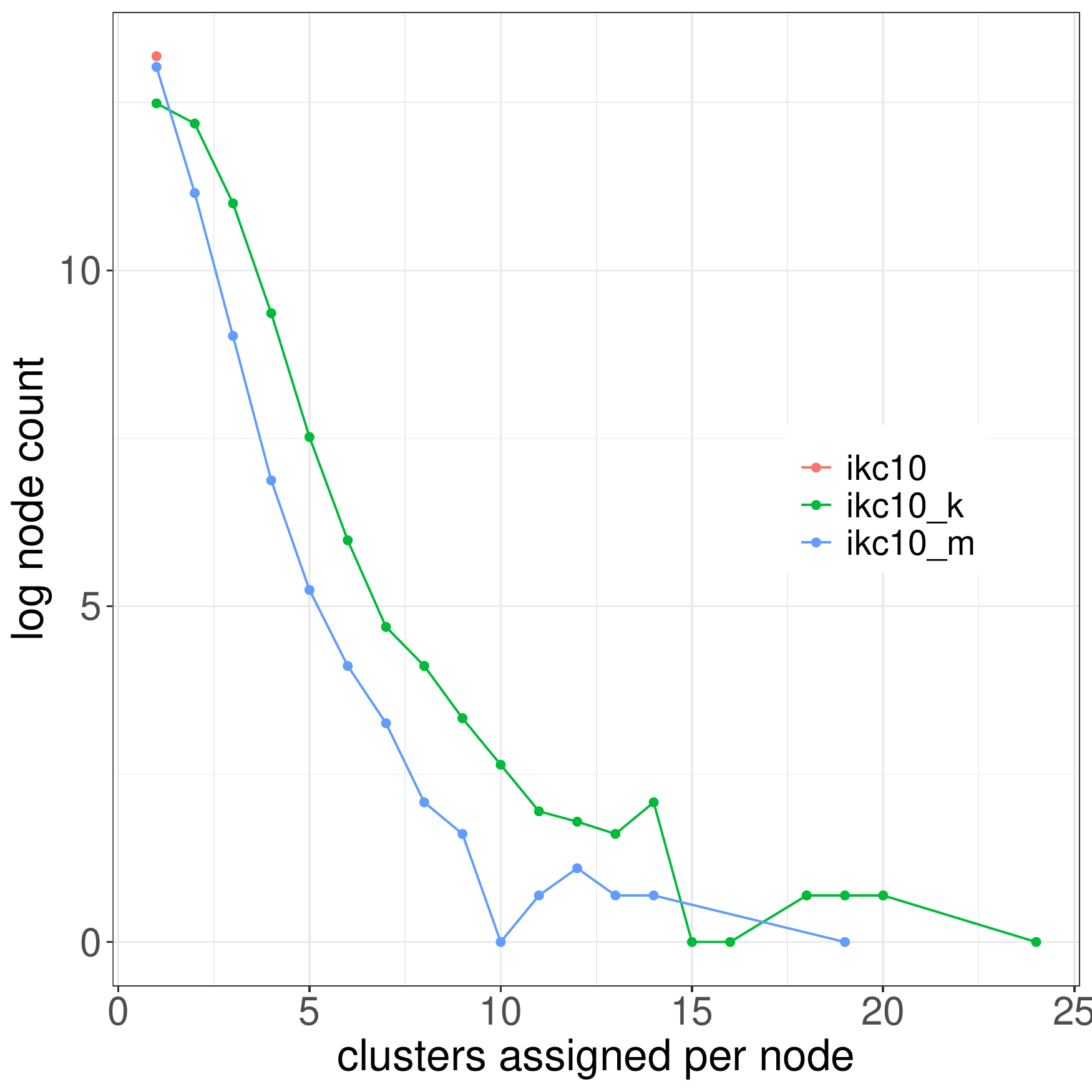} 
\end{subfigure}
\hfill
\begin{subfigure}[t]{0.48\textwidth}
\centering
\includegraphics[width=\linewidth]{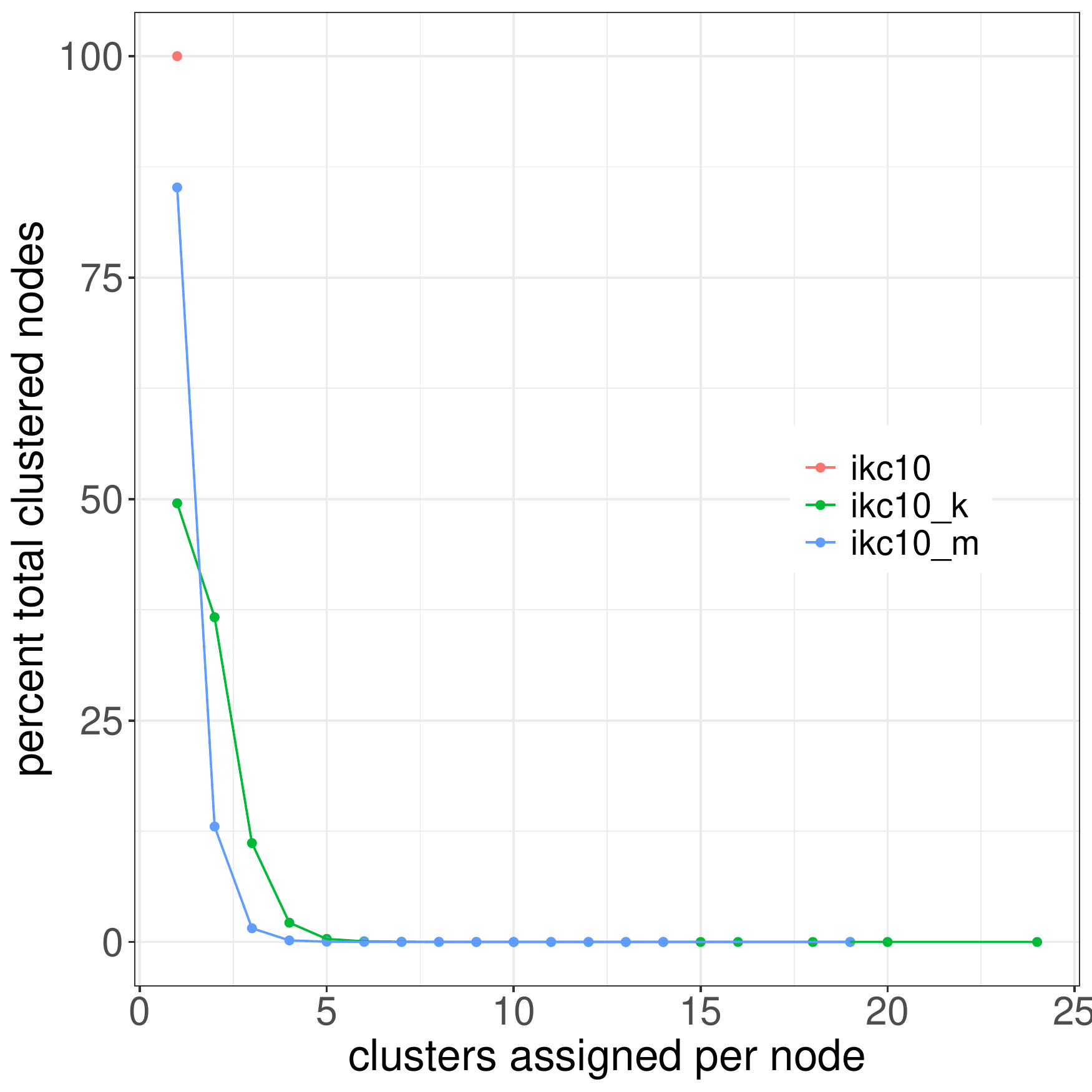} 
\end{subfigure}
\captionsetup{width=0.9\textwidth}	
\caption{AOC selectively assigns nodes to multiple clusters. The count of nodes plotted against how many clusters a node was assigned to after AOC treatment enforcing either \emph{k} (ikc10\_k, green) or \emph{mcd} (ikc10\_m, blue);  these are shown as natural log counts (left panel) or percentages of the number of nodes (right panel) in non-singleton clusters. One node is assigned to 24 different clusters in the case of AOC\_k. The single red point in both plots (top left) indicates that all the 535,165 nodes in the input IKC\_k10 clustering are in single clusters. Cluster size is shown on the y-axis in natural log units. Log values of 5 and 10, and 15 correspond, after rounding, to 148 and 22,026 respectively.}
\label{fig:fig2}
\end{figure}

We then examined the number of clusters a candidate node was assigned to after AOC treatment of clusterings generated by  IKC. For AOC\_k treatment of IKC\_k10 clusters, 54\% of nodes in non-singleton clusters were assigned to between 2 and 24 clusters in a progressively decreasing manner, with roughly 26\% of nodes assigned to 2 clusters and a single node being assigned to 24 different clusters.  The remaining 46\% of the nodes were assigned to a single cluster. AOC\_m, in comparison to AOC\_k, results in fewer multiple cluster assignments because of its more stringent membership criterion (Figure~\ref{fig:fig2}).
	
\subsection{Experiment 3: Which node properties impact multiple assignments?}
\begin{figure}
\centering
\begin{subfigure}[t]{0.48\textwidth}
\centering
\includegraphics[width=\linewidth]{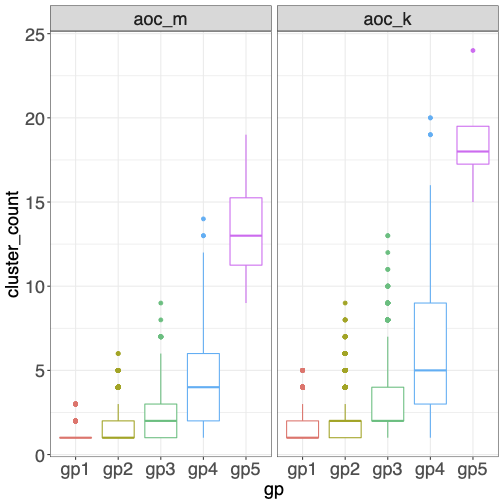} 
\end{subfigure}
\hfill
\begin{subfigure}[t]{0.48\textwidth}
\centering
\includegraphics[width=\linewidth]{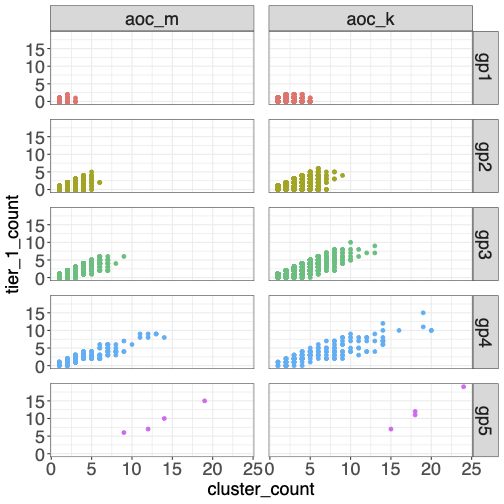} 
\end{subfigure}
\captionsetup{width=0.9\textwidth}	
\caption{Cluster and tier assignments by node degree. The subfigure on the left shows the distribution of numbers of clusters each publication belongs to, and the subfigure on the right shows cluster count by Tier 1 membership for each group; each subfigure presents these results for both AOC\_m and AOC\_k. Nodes are partitioned into five groups based on their total in-network degree (in\_degree + out\_degree) with group 1 containing the nodes with the smallest total degree and group 5 with the nodes with the largest total degree. Group 1 has the largest number of observations and group 5 has the smallest. Group statistics (group; class limit; number of nodes): [gp1; $<$100;  272,395], [gp2; 100-999; 254,832], [gp3; 1,000-9,999; 7,773], [gp4; 10,000-99,999; 161], and [gp5; $\geq$100,000; 4]. }
\label{fig:fig3}
\end{figure}
	
To ask whether node degree is associated with the number of clusters the node is assigned to by AOC, we examined clustered nodes in the CEN and compared their total in-graph degree to the number of clusters they were assigned for either in AOC\_m or AOC\_k treatment of IKC\_k10 clustering of the CEN (Figure~\ref{fig:fig3}). We partitioned the nodes into five groups based on their total (in\_degree + out\_degree) in-network degree ,  with group 1 containing the nodes with the smallest total degree (less than 100), group 2 with nodes of total degree between 100 and 999, group 3 with nodes of total degree between 1,000 and 9,999, group 4 with nodes of total degree between 10,000 and 99,999, and group 5 the nodes with total degree at least 100,000.

Over 98\% of these 535,165 nodes are in groups 1 and 2; 50.9\% in group 1, 47.6\% in group  2, 1.5\% in group 3,  0.03\% in group 4, and 0.0007\% in group 5. 
Although group classification is based on total degree,  any publication in groups 2-5 with high total degree must have high in-degree (in-network citation count). Membership in group 3  reflects a large citation count (at least 750 in-graph citations), and membership in groups 4 or 5, the top 169 nodes by total degree, reflects ultra-large citation counts (at least 9,750 in-graph citations).    
	
We first compare clusterings produced by AOC\_m and AOC\_k (Figure~\ref{fig:fig3}, left subfigure). For both AOC\_m and AOC\_k, the number of clusters that any node is assigned to increases as we move from group 1 to group 5, showing that, in general, total degree is associated with the number of clusters that a node is assigned to. The number of cluster assignments per node is larger for AOC\_k than for AOC\_m, which is not unexpected since AOC\_m is a more stringent membership criterion. Hence, in-network citation count is associated with the number of clusters that a publication is a core member of.
	
However, for both AOC\_k and AOC\_m, the distributions for each group are overlapping, revealing potentially interesting differences between publications that are not explained just by citation count. 
Examining AOC\_k, for example, we see the following trends. The largest number of clusters any node is assigned to is 25 and the smallest number is 1. All the nodes assigned to 14 or more communities are in groups 4 or 5, and so have total degree at least 10,000. In addition, group 5 publications belong to a minimum of 14 clusters. 

In contrast, every other group has publications that belong to only 1 cluster. The largest number of communities for publications in groups 1 and 2 is  9,  group 3 publications appear in at most 13 communities, group 4 publications appear in up to 20 communities, and group 5 publications appear in up to 25 communities.  Since only group 5 publications appear in more than 20 communities, we conclude that, under the conditions of our clustering, an ultra-high in-graph citation count is  necessary  for assignment to a large number of communities.  Results for AOC\_m are similar but with reductions in the total number of clusters each publication can be in,  which follows because AOC\_m is a more restrictive condition than AOC\_k.

While there appears to be an association between the degree of a node and the likelihood of it being assigned to multiple clusters, there are instances where nodes of high degree are assigned to only one or two clusters. 
For example, for AOC\_k, five publications are found in Group 4 that are assigned to only one cluster \citep{arnon1949copper,ellman1961new,friedewald1972estimation,iijima1991helical,raymond1995genepop}. Four of these five describe methods. All five were published in or before 1995 (1949-1995), have high in-degree (12,741 to 32,927) and very low out-degree in our data (0-11). Whether this low out-degree contributed to restricted cluster assignment merits follow up and opens up the questions of breadth and dependence \citep{bu2021multidimensional} in publication communities, as well as that of data quality. This shows that high total degree is not sufficient for membership in many communities and underscores the case for mixed methods approaches.
	
Another perspective that provides additional insight into publications is their ``tier" within their communities. In \cite{Chandrasekharan2021}, we  proposed a tier classification for nodes in a cluster, in which Tier 1 refers to the nodes in the top 10th percentile with respect to intra-cluster citations. Thus, when measuring in-degree within the cluster, a Tier 1 node is in the top 10 percent compared to all other nodes in its cluster. 

We observe that nodes assigned to multiple clusters are more likely to have Tier 1 status (Figure~\ref{fig:fig3}, right subfigure). The Tier 1 count also increases for AOC\_k compared to AOC\_m. 
The greater Tier 1 count for AOC\_k is likely a combination of larger clusters and the in-degree of nodes within them.

For groups 1--4, there are some publications that are Tier 1 in all communities they belong to, some that are never in Tier 1, but the majority are in between. 
However, while the four publications in group 5 are Tier 1 for at least 7 communities for AOC\_k, only one is in Tier 1 for more than 10 communities. 

Interestingly, under AOC\_k there are four publications in Group 4 that are in Tier 1 for at least 10 communities, two that are Tier 1 in strictly more than 10 communities, and one of these is Tier 1 in 15 communities. Also in AOC\_k, we find publications in group 2 that are Tier 1 in up to 7 communities, publications in group 3 that are Tier 1 in 8 or more communities, and a publication in group 4 that is in Tier 1 for 15 communities. Results under AOC\_m also show similar trends but with lower total Tier 1 counts, consistent with the reduced number of clusters that each publication belongs to.
	
These trends show that while total degree is correlated with the number of clusters a publication belongs to and how many clusters it is in Tier 1 for,
these values are not determined just by total degree. Thus, the cluster membership and tier status within their communities provides complementary insights into  the publications that goes beyond citation count.
	
\subsection{High Degree Singleton Nodes} 
	
IKC clustering of the CEN results in 3.8\% coverage. A large number of nodes of high in-network degree are assigned to singleton clusters and not to cores. In the case of the CEN clustered by IKC\_k10, 15,039 nodes in the top 1\% (by degree) of nodes in the network are assigned to singleton clusters. We examine here whether this population can be reduced with AOC.

\begin{figure}
\centering
\includegraphics[width=0.6\linewidth]{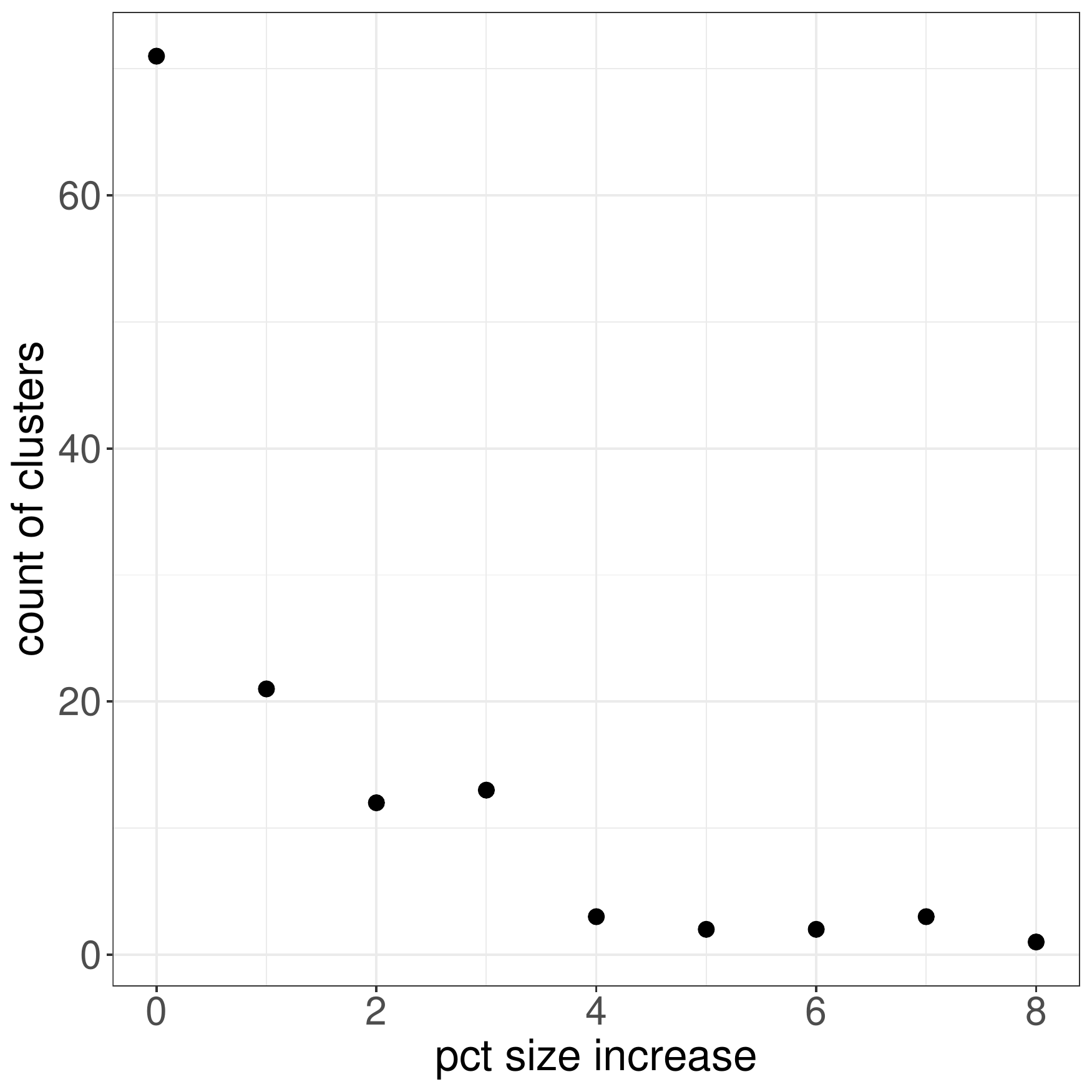} 
\captionsetup{width=0.9\textwidth}
\caption{AOC\_k but not AOC\_m incorporates high-degree singleton nodes from IKC\_k10 into clusters. 15,039 nodes belonging to the top  1\% of nodes by total degree in the CEN were assigned to singleton clusters after IKC\_k10 clustering. These 15,039  ``singleton" nodes were used as candidate nodes for AOC\_m and AOC\_k treatment of IKC\_k10 clusters. After AOC\_m treatment, none of these 15,039 nodes were incorporated into any of the 128 clusters resulting from  IKC\_k10. After the more permissive AOC\_k protocol, 7,459 (49.6\%) were incorporated into one or more of the 128 IKC\_k10 clusters. The count of clusters is plotted against percent increase in cluster size for all 128 clusters. 65 of 128 clusters did not change in size. Of the remaining 63 clusters, the maximum increase in cluster size was 8.1\%.  After AOC\_k treatment, 540,883 nodes (99\%) of the nodes were assigned to one cluster, 1504 to 2 clusters, 210 to 3 clusters, and 26 to 4 clusters; one node was assigned to 5 clusters.
\label{fig:singleton}}
\end{figure}

With AOC\_m using these 15,039  ``singleton" nodes as candidates, no additional assignments are made and so the output of AOC\_m is identical to IKC\_k10. 
With the lower stringency AOC\_k, however, 7,459 of 15,039 (49.6\%) of the singleton nodes are assigned to IKC\_k10 clusters, with all nodes assigned to at most 5 clusters, most nodes assigned to only one cluster, and only one node assigned to 5 clusters (Figure~\ref{fig:singleton}). 

In terms of cluster size increases, the effect is also mild: 51\% of the 128 clusters in this AOC\_k treatment do not increase in size. In contrast, when the candidates are nodes that are not singletons (so that they belong to a non-singleton cluster), 74\% and 87\% of the 128 clusters increase in size with AOC\_m and AOC\_k respectively  (Section 3.2).  

Thus, AOC enables nodes previously assigned to singleton clusters to  be incorporated into cores, and while these assignments may not impact the clusters significantly, these assignments may shed light into the roles of these publications within the network.  Whether this option is useful will depend on the purpose of clustering and the evaluation criteria designed by users for a specific study.

\subsection{Experiment 4: Marker node concentrations}

\begin{figure}[H]
\centering
\includegraphics[width=0.7\linewidth]{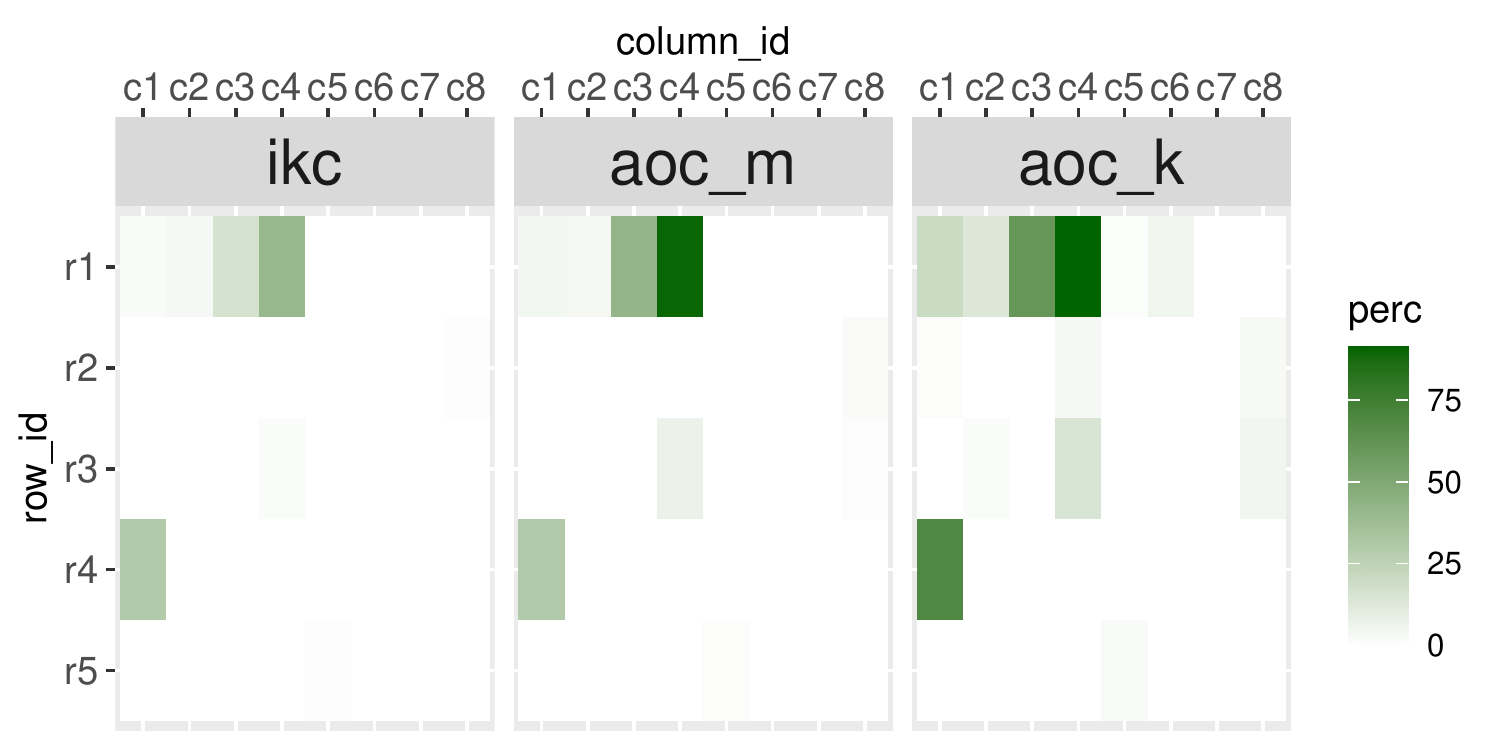} 
\captionsetup{width=0.9\textwidth}
\caption{Marker Node Enrichment with AOC. We show marker node counts in 40 clusters (5 rows with 8 clusters per row) before and after AOC.  The count of clusters with non-zero marker node counts is maximal in the case of AOC\_k (right panel), with 31 clusters containing markers. Notably, the proportion of 1,021 marker nodes in the network increases from 40.7\% in cluster 4 (r1,c4) of IKC clustering to 90.2\% after AOC\_m to 91.3\% after AOC\_k. The proportion of markers in cluster 25 (r4,c1) is the same (30.4\%) for IKC and AOC\_m but increases to  69.5\% under the more permissive conditions of AOC\_m. Data are shown for clusters where 1\% or more of the markers are present in any of IKC, AOC\_m, or AOC\_k. \emph{Perc}: Percentage of 1,021 marker nodes found in a cluster.}
\label{fig:marker-node-concentration}
\end{figure}

In the preceding sections, we examined the effects of AOC from a graph-theoretic one aimed at generalizability. We now introduce a contextual perspective, in which we interpret findings relative to a field of interest. In this case, we are studying communities in the field of extracellular vesicle (EV) research. Context is addressed in two ways. We use, as input to IKC, a citation network (CEN) that is enriched in the recent extracellular vesicle literature. IKC reduces this large network to 128 cores that contain 3.8\% of the nodes in the network. From these 128 cores, we identify a subset of interest by using a set of markers derived from the cited references of recent reviews of the extracellular vesicle field that were authored by different researchers \citep{Wedell2022}. Under the assumption that cores enriched in marker nodes are relevant to extracellular vesicle research, we further reduce the data under consideration to those cores. We now assess how AOC impacts marker node concentration (Figure~\ref{fig:marker-node-concentration}).

The count of cores with non-zero marker counts varied between treatments. For IKC, 17 of 128 cores exhibited  non-zero marker counts. Clusters 3, 4, and 25 are notable in accounting for 87.5\% of 1021 markers after IKC clustering. Because of this substantial coverage of marker nodes, from the perspective of EV biology, clusters 3, 4, and 25 are of obvious interest and offer a significant reduction in the amount of information to be studied qualitatively.  

After IKC+AOC treatment, 20 of 128 (AOC\_m) and 31 of 128 (AOC\_k) cores respectively contained non-zero marker counts, which is consistent with their relatively stringent and permissive design. After AOC\_m treatment of IKC clusters, clusters 3, 4, and 25  contained 42.5\%, 90.2\%, and 30.4\% respectively, of all markers. After AOC\_k treatment of IKC clusters, clusters  3, 4, and 25 contained 60.5\%, 91.3\%, and 69.5\% of all markers respectively.  Thus,  clusters 3, 4, and 25 all expand more significantly under AOC\_k than under AOC\_m. We also note that cluster 1, which contains 2.4\% of markers in IKC and 4.8\% after AOC\_m, is significantly enriched for marker nodes by AOC\_k to 20.7\% of the marker nodes. These data  suggest that the recursive approach of IKC results in markers being segregated by disjoint clustering, but this effect can be remediated by post-processing using AOC.

\clearpage
	
\subsection{Examining overlap between clusters}

The use of marker nodes is one approach to identify clusters of relevance. After enrichment by AOC\_k or AOC\_m, clusters will overlap, and the overlap consists of 
a mixture of marker and non-marker nodes.   

Accordingly, we examined overlap between clusters after AOC\_m or AOC\_k treatment of IKC clusters, and we specifically observe relationships between clusters 3, 4, and 25, which were previously shown to be rich in marker nodes. Weighted edges were drawn between clusters based on the Jaccard Coefficient (ratio of intersection/union) for overlap. A threshold of the median Jaccard Coefficient from all values was set to permit an edge. Because some clusters do not have sufficient intersection with any others  they do not have any incident edges in these graphs.

\begin{figure}[H]
\centering
\begin{subfigure}[t]{0.48\textwidth}
\centering
\includegraphics[width=\linewidth]{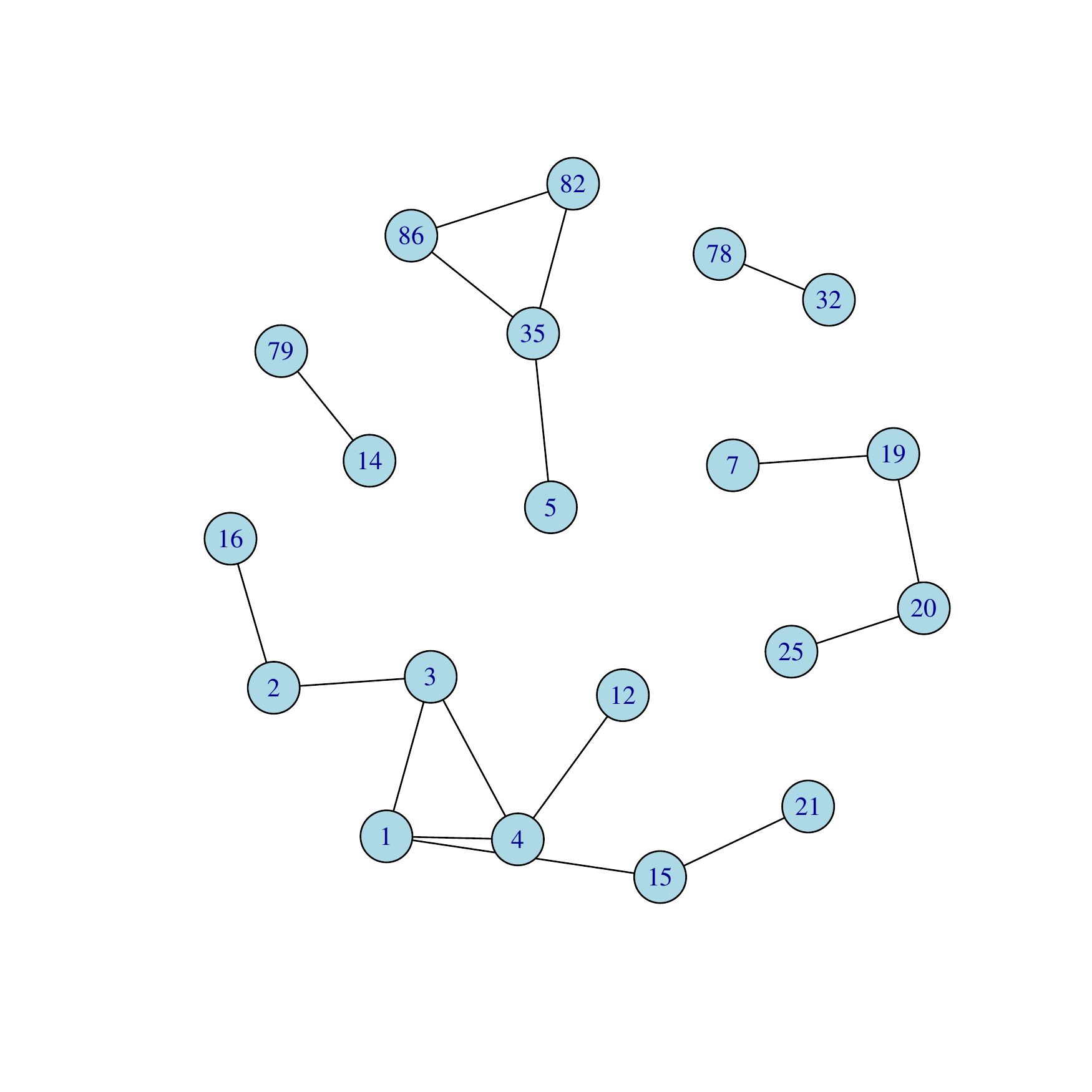}
\caption{AOC\_m}
\end{subfigure}
\hfill
\begin{subfigure}[t]{0.48\textwidth}
\centering
\includegraphics[width=\linewidth]{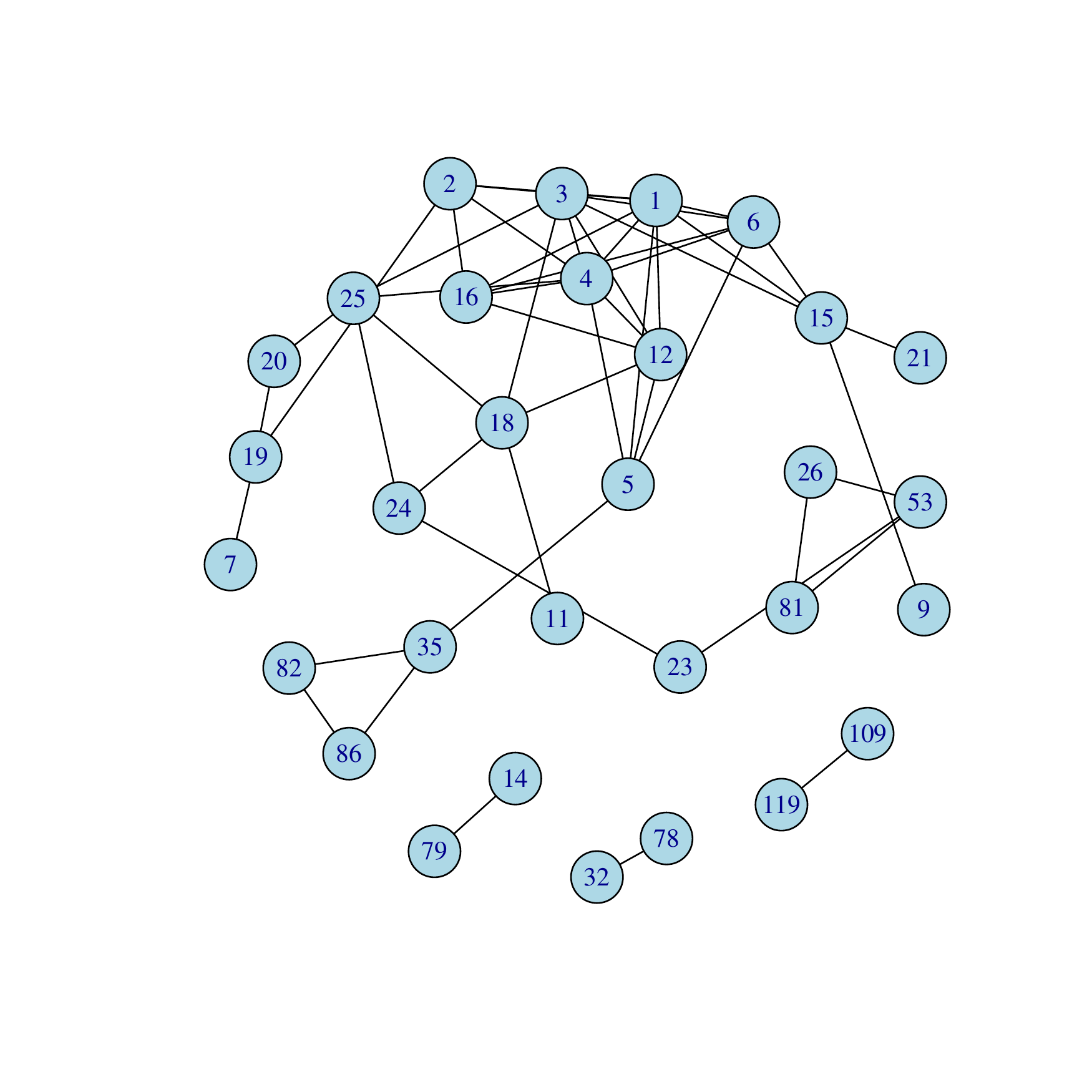} 
\caption{AOC\_k}
\end{subfigure}
\caption{Overlapping clusters produced by IKC\_k10 + AOC.  Overlapping clusters were generated from CEN data using IKC\_k10 followed by either treatment with either AOC\_m (left) or AOC\_k (right). The set of candidate nodes presented to AOC was all nodes in non-singleton IKC clusters. Edges are drawn between clusters based on the Jaccard Coefficient for node overlap and are visible if the JC exceeds the median value for all pairs. Cluster numbers in both panels correspond to cluster numbers from the input IKC clustering. Clusters 3, 4, and 25 are enriched for markers.  (Clusters 4 and 25 are adjacent in IKC\_k10+AOC\_k, although the edge connecting them passes through node 16 in the visualization.)}
\label{fig:overlapping}
\end{figure}

These networks, shown in Figure \ref{fig:overlapping} (left shows the result for AOC\_m and right shows the result for AOC\_k), contain only those clusters that have at least
one edge. Note that  the left network (AOC\_m) has fewer nodes than the right network (AOC\_k) which indicates that augmentation using AOC\_m does not produce substantial 
overlap for as many clusters as augmentation by AOC\_k.

Beginning with the AOC\_m network, we note that the network forms five connected components, with clusters 3 and 4  in the same component and connected by an edge.
In contrast, cluster 25 is in a separate component. Thus, clusters 3 and 4 identified in IKC have substantial overlap after AOC\_m, indicating some shared research questions or approaches, but
cluster 25 reflects a different population. 

Turning to IKC+AOC\_k,  as noted earlier, we see a larger number of clusters, indicating that more of the clusters had sufficient overlap with other clusters to be retained in the network visualization. 
Interestingly here we see that clusters 3, 4, and 25 are all pairwise connected by edges, indicating that all three have a significant number of shared publications after AOC\_k. 
Hence,  what we see here is that the overlap after AOC\_m between cluster 25 and the other two clusters is too weak to be considered significant, but is sufficient  after AOC\_k to be considered significant.
Since AOC\_m represents a more restrictive criterion, this shows that clusters 3 and 4 have a strong relationship  to each other and a somewhat weaker relationship to cluster 25 but one that is nevertheless worth noting.

While both AOC\_k and AOC\_m offer insights into the EV research community structure,   AOC\_m provides more specific information about the research communities, since it is a local constraint that preserves the MCD of each core, compared to AOC\_k which only enforces the value of $k$ across all the cores and may therefore add weakly linked nodes to the IKC cores. 
On the other hand, the examination of clusters 3, 4, and 25 given above shows that by combining information obtained from both AOC\_m and AOC\_k, additional insights can be obtained.  
Therefore, we allow both options and allow users to choose between them.

\clearpage
	
\section{Conclusions} We developed AOC as a meta-method for overlapping clusters that serves as an option for users of the IKC pipeline. \added[id=gc]{The overarching vision for this pipeline is ambitious: a scalable modular workflow that supports identifying and characterizing research communities. We are positioned between method development and exploratory discovery. 

The pipeline begins with constructing a citation network. It ends, after traversing multiple automated stages, with interpretation by experts. The work described in this article reflects focus on the narrower question of extending IKC by using AOC to inclusively identify the cores of core-periphery communities.}

We sought to offer multiple options to users. This is achievable through varying the input data, the $k$ setting for IKC, the two AOC  options for membership, and the choice of candidate nodes. \added[id=gc]{To test AOC, we use a citation network centered around the recent extracellular vesicle literature, a rapidly growing field in biology \citep{van2022challenges}.} 

\added{The study has two complementary objectives: (i) to enhance a modular community finding pipeline that is relatively subject-independent and (ii) examine its effects on clustering in a case study of the extracellular vesicle literature.}

For the first, we have identified citation-dense communities of publications using IKC, a k-core based approach. We have enriched these clusters by applying two variants of AOC. In this respect, we are able to address a limitation of our original IKC method that arises by its restriction to producing disjoint clustering (as a result of extracting k-cores in decreasing order of the value of $k$). This limitation prevents a node captured in a k-core from being considered for inclusion in a subsequently extracted k-core. Post-processing with AOC overcomes this limitation.

Since membership in multiple clusters following AOC occurs for many nodes, and is not completely predictable based on degree within the network, a benefit of AOC is that examination of the clusters a publication belongs to, and the role of the publication within these clusters,  may provide additional insights into the roles of publications within the network that go beyond evaluation based on citation count.
Such studies require expertise in the disciplines for the publications, and thus provides opportunities for specialists for future investigation.

For the second, a study of the extracellular vesicle literature, we have sought to include human experience and intent \citep{vonluxburg2012clustering} in AOC through controlling the input data and enabling contextual evaluation in the form of externally identified markers. The results with the CEN suggest that AOC tends to enrich those clusters, already rich in markers. On the one hand, this property may not be very useful in identifying the most marker-dense clusters but it does provide a more complete description for follow-on studies. 

Finally, we note that AOC could be applied to clustering  outputs from algorithms other than IKC. Further, it could also be engineered to accommodate new membership criteria such as average cluster degree.  discovery perspective, This is the subject of future work. 
	
\section*{Competing Interests} \vspace{3mm} The authors have no competing interests. 
	
\section*{Funding Information} TW receives funding from the Grainger Foundation. Research reported in this manuscript was supported by the Google Cloud Research Credits program through award GCP19980904 to GC.
	
\section*{Data Availability} Access to the bibliographic data analyzed in this study requires access from Digital Science. Code generated for this study is freely available from our Github site \citep{Liu_BL_2021}. Supplementary
materials are available at the same location. Retraction data used to curate the networkk are available from The Center For Scientific Integrity, the parent nonprofit organization of Retraction Watch, subject to a standard data use agreement. Dimensions data were made available by Digital Science through the free data access for scientometrics research projects program.
	
\section*{Acknowledgments} AJ is presently in the graduate program at Princeton University; his contributions to this manuscript were made while he was a computer science major at the University of Illinois Urbana-Champaign.
 We thank Srijan Sengupta from North Carolina State University for critical advice. We thank Alison Abritis and Ivan Oransky from Retraction Watch for helpful suggestions and for making data available. We thank Digital Science, Google, and the Grainger Foundation. 

\bibliographystyle{apalike}
\bibliography{aocR1}
\end{document}